\documentstyle[12pt]{article}
\newcommand{\be}{\begin{equation}}
\newcommand{\ee}{\end{equation}}
\newcommand{\ba}{\begin{eqnarray}}
\newcommand{\ea}{\end{eqnarray}}
\newcommand{\ban}{\begin{eqnarray*}}
\newcommand{\ean}{\end{eqnarray*}}

\newcommand{\n}[1]{\label{#1}}

\def\bbox{{\,\lower0.9pt\vbox{\hrule \hbox{\vrule height 0.2 cm
\hskip 0.2 cm
\vrule  height 0.2 cm}\hrule}\,}}

\def\bbox{{\,\lower0.9pt\vbox{\hrule \hbox{\vrule height 0.2 cm
\hskip 0.2 cm
\vrule  height 0.2 cm}\hrule}\,}}

\begin{document}
\setlength{\unitlength}{1mm}

\title{{\hfill {\small Alberta-Thy-05-04, \ MCTP-04-20 } } \vspace*{2cm}
 \\
Interaction of higher-dimensional
rotating black holes with branes}
\author{
Valeri P. Frolov\footnote{e-mail: frolov@phys.ualberta.ca}$~{}^1$,
Dmitri V. Fursaev\footnote{e-mail: fursaev@thsun1.jinr.ru}$~{}^2$,
 and
Dejan Stojkovi\'{c}\footnote{e-mail: dejans@umich.edu}$~{}^3$
}

\date{}
\maketitle
\noindent
\centerline{$^{1}$\em
Theoretical Physics Institute, Department of Physics,}
\centerline{ \em University of Alberta,  Edmonton, Canada T6G 2J1}
\\ 
\centerline{$^{2}${\em Joint Institute for
Nuclear Research,}}
\centerline{\em Bogoliubov Laboratory of Theoretical Physics,}
\centerline{\em  141 980 Dubna, Russia}\\
\centerline{$^{3}${\em MCTP, Department of Physics, University of
 Michigan,}}
\centerline{\em  Ann Arbor, MI 48109 USA}
\bigskip

\noindent

\begin{abstract}   
We study interaction of rotating higher dimensional black holes with
a brane in  space-times with large extra dimensions.  We demonstrate
that in a general case a rotating black hole attached to a brane can
loose bulk components of its angular momenta. A stationary black hole
can have only those components of the angular momenta which are
connected with Killing vectors generating transformations preserving
a position of the brane. In a final  stationary state  the null
Killing vector generating the black hole  horizon is tangent to the
brane.  We discuss first the interaction of a cosmic string and a
domain wall with the 4D Kerr black hole. We then prove the general
result for slowly rotating higher dimensional black holes interacting
with branes. The characteristic time when a rotating black hole with
the gravitational radius  $r_0$ reaches this final stationary state
is $T\sim r_0^{p-1}/(G\sigma)$, where $G$ is the higher dimensional
gravitational coupling constant, $\sigma$ is the brane tension, and
$p$ is the number of extra dimensions.   
\end{abstract}

\vspace{.3cm}

\newpage

%%%%%%%%%%%%%%%%%%%%%%%%%%%%%%%%%%%%%%%%%%%%%%%%%%%%%%%%%%%%%%%%%%%%%%%%%%
%%

\baselineskip=.6cm
\section{Introduction}

Models  \cite{ADD,RS} where a space-time has large or infinite extra
dimensions became very popular recently. In these models the
fundamental energy scale can be as low as a few TeV which opens a
possibility of the experimental tests of predictions of these models
in the future collider and cosmic ray experiments \cite{EG,Land:03}. 
In a higher dimensional space (with a number of extra dimensions $p$)
at distances $r$ much smaller than the size of extra dimensions,  the
gravitational force $GM/r^{1+p}$ becomes much stronger than in 4D
space. For this reason in the interaction of particles with  TeV
energies one can expect considerable emission of bulk gravitons. This
effect could be observed as the energy-momentum non-conservation
in the brane world. Another dramatic prediction of these models is a
possibility of creation of mini black holes when the center of mass
energy of two colliding particles becomes higher than a fundamental
energy \cite{EG}. 

Being gravitational solitons black holes can propagate in the bulk
space and thus  serve as probes of extra dimensions. We
focus our attention on black holes with a size much smaller than the
size of extra dimensions. In this case the effects  of the
extra dimensions on the black hole geometry are small and can be
neglected. (For recent review of black holes in a space-time with
large extra dimensions see \cite{kanti}.)

In a general case a black hole created by a collision of two
particles is rotating.  If there is no emission of the bulk
gravitons   the black hole rotation occurs within the brane. In a
more realistic situation when  bulk gravitons are emitted, a black
hole can also acquire bulk components of the angular momentum. There
are also other processes which may result in the black hole  rotation in
the bulk dimensions. For example,  if a black hole collides with a
particle or another black hole with emission of bulk gravitons, or
when it emits bulk gravitons in the Hawking evaporation process.

In the approximation when the tension of the brane is  neglected and
a black hole gravitational radius is much smaller than the size  of
extra dimensions, the higher dimensional rotating black holes are
described by the Myers-Perry (MP) metric \cite{MP}\footnote{It should
be emphasized that in the higher dimensional case a black-hole's
horizon can has the topology which differs from the topology of the
sphere\cite{EmRe:02} and rapidly rotating black holes can be unstable
\cite{EmMy:03}. For simplicity we do not consider these effects.}. In
this paper we discuss the effects of interaction of a rotating higher
dimensional black hole with a brane. We shall make two simplifying
assumptions: (i) The brane--black-hole interaction is weak; (ii) A
black hole is slowly rotating. The first assumption means that we
take into account corrections linear in terms of the brane tension.
The second assumption implies that we shall study only linear in the
rotation parameter corrections. In this approach, we can start with a
static unperturbed solution for a test brane in a space-time of a
non-rotating black hole (section~2) and consider the perturbation of
the brane induced by a small rotation of the black hole. There exists
an evident static solution of Nambu-Goto equations describing a
static thin brane, namely a configuration when the world-sheet
representing the brane coincides with the `equatorial plane' of the
black hole. The equatorial plane of a spherically symmetric space-time
is invariant under a reflection transforming the northern hemisphere
into the southern one and thus it is a geodesic sub-manifold. Since
any geodesic surface is also a minimal one, this equatorial plane
obeys the Nambu-Goto equations.

We start our study of interaction of slowly rotating black holes with
branes by considering the simplest example in 4D space, namely a
thin domain wall interacting with the Kerr black hole (section~3).
According to a well known result proved by Hawking in 1972
\cite{Hawk:72} a stationary black hole either does not rotate and its
metric is static, or it rotates and the space-time possesses
additional axial symmetry.  In the latter case the matter
distribution around stationary rotating black hole   must be axially
symmetric, which means that the Lie derivative of the matter
stress-energy tensor $T^{\mu\nu}$ along the Killing vector field
$\xi_{\varphi}$ generating rotations of the black hole is vanishing,
${\cal L}_{\xi_{\varphi}}T^{\mu\nu}=0$. This result implies that if
there exists a matter distribution violating axial symmetry its
gravitational interaction with the black hole slows down its rotation
until it becomes static.

If a black hole has a rotation around the axis orthogonal to the
domain wall, the configuration is axisymmetric and that there is no
angular momentum transfer from the black hole to the domain wall in
this case. A situation is quite different when the axis of rotation
is not orthogonal to the unperturbed domain wall. A system looses its
axial symmetry and one can expect  the angular momentum transfer from
a black hole to the brane. In  section~3 we calculate  the rate of
the angular momentum flux.

It is interesting that in the case of a cosmic string, a solution for
stationary test string attached to the black hole is known exactly
for an arbitrary value of the rotation parameter. Using this solution
it is possible to demonstrate that a rotating black hole attached to the
string looses its angular momentum. We calculate the rate of the
angular momentum loss. This process continues until the system
reaches a final stationary state where the string coincides with the
axis of rotation of the black hole. These results are presented in
the Appendix.

In the second part of the paper we discuss the interaction 
of  rotating black holes  with  branes in a space-time
with large extra dimensions. In many aspects this generalization is
straightforward. One only should remember that in a highter
dimensional case there are several mutually commutative and
orthogonal space-like Killing vectors $\xi_{(i)}$ singled out by the
property that they have closed integral lines. The rotation of the
black  hole is characterized by a set of angular velocities
$\Omega_i$ of the horizon along corresponding integral lines of
$\xi_{(i)}$. We demonstrate that  a slowly rotating
higher-dimensional black hole interacting with a static brane 
looses some of the components of its angular momenta and reaches a
final stationary equilibrium state. In this state the stress energy
tensor of the brane satisfies the following condition $\sum_i
\Omega_i {\cal L}_{\xi_{(i)}}T^{\mu\nu}=0$. The final  state
also possesses the property that the null Killing vector generating
the black hole horizon is tangent to the brane, which means that
there is no energy fluxes through the black hole horizon.

Our calculations show that the characteristic time of the relaxation
during which a rotating black hole reaches the equilibrium state is
shorter than the time during which it looses its bulk angular momentum
because of the Hawking radiation. This may have important experimental
signature of mini-black holes in future collider and
cosmic ray experiments.

We conclude the paper by considering special cases of mutual
orientation of a rotating black hole and a brane and illustrate the
general conclusions (section~6). In the discussion (section~7) we
comment on brane--black-hole interaction in the Randall-Sundrum brane
worlds.

\section{A black hole attached to a brane}
\setcounter{equation}0

Before discussing the effects connected with the rotation of black
holes, we consider a simpler case of static brane attached to a
non-rotating black hole. In the approximation when the  gravitational
back-reaction of the brane is neglected its world-sheet obeys the
Nambu-Goto equation
\be\n{1}
^{(n+1)}\Delta X^{\mu}+\gamma^{ab}\Gamma^{\mu}_{\ \
\nu\lambda}X^{\nu}_{,a}X^{\lambda}_{,b}=0~~.
\ee
The relations $X^{\mu}=X^{\mu}(\zeta^a)$ determine the embedding of
the $n+1$-dimensional brane into $N+1$-dimensional  bulk space-time.
$\zeta^a$, ($a,b=0,n$), are internal coordinates in the brane and
$X^\mu$, ($\mu=0,N$) are  coordinates in the bulk space with the
metric $g_{\mu\nu}$. The connections $\Gamma^{\mu}_{\ \ \nu\lambda}$
are determined for  $g_{\mu\nu}$.
To exclude degenerate cases we assume that
$0<n<N$. $^{(n+1)}\Delta$ is the box-operator for the
induced metric
\be\n{2}
\gamma_{ab}=g_{\mu\nu}\, X^{\mu}_{,a}X^{\nu}_{,b}\, .
\ee
The stress-energy  tensor of the brane is
defined as follows:
\be\n{3}
\sqrt{-g}T^{\mu\nu}=-\sigma \int d^{n+1}\zeta \delta^{(N+1)}(X-X(\zeta))
\sqrt{-\gamma} \gamma^{ab}X^{\mu}_{,a}X^{\nu}_{,b}~~\, , 
\ee
where $\sigma$ is the tension of the brane.

The metric Tangherlini  of a non-rotating higher dimensional black
hole is  \cite{Tang}
\be\n{4}
d\bar{s}^2=-Bdv^2+2drdv+r^2d\Omega^2_{N-1}\, ,\hspace{0.2cm}
B=1-\left({r_0 \over r}\right)^{N-2}~~.
\ee
The coordinate $v$ is the advanced time and $d\Omega^2_{N-1}$ is the
metric on the unit sphere $S^{N-1}$. For $N=3$ this metric reduces to
the Schwarzschild metric. The gravitational radius $r_0$ is related to
the black hole mass $M$ as follows
\be\n{mass}
M={(N-1){\cal A}_{N-1}\over 16\pi G_{N+1}} r_0^{N-2} \, ,
\ee
where
\be\n{vol}
{\cal A}_{N-1}={2\pi^{N/2}\over \Gamma(N/2)}
\ee 
is the area of a
unit sphere $S^{N-1}$ and $G_{N+1}$  is the
$N+1$-dimensional gravitational coupling constant which has
dimensionality $[mass]/[length]^{(N-2)}$.

Consider a unit sphere $S^{N-1}$ embedded in a $N$-dimensional
Euclidean space $R^N$, and let $X^A$, ($A=1,\ldots,N$) be the
Cartesian coordinates in $R^N$. One can choose these coordinates so
that the equations $X^{n+1}= \ldots = X^N=0$ determine the
$n$-dimensional hyper-surface (brane). This hyper-surface intersects
the unit sphere $S^{N-1}$ along a surface ${\cal S}$ which has a
geometry of a round unit sphere $S^{n-1}$. The surface ${\cal S}$ is
a higher dimensional analogue of a `large circle' on a
two-dimensional sphere.  In particular, being considered as a
sub-manifold of $S^{N-1}$ it has a vanishing extrinsic curvatures, and
hence is a geodesic sub-manifold. We denote by $\omega^{\alpha}$
coordinates on ${\cal S}$, and by $d\omega_{n-1}^2$ the metric on it.

One can construct a solution for a static $n$-brane as follows. We
use $\zeta^{a}=(v,r,\omega^{\alpha})$ as coordinates on the brane. Then 
\be\n{7}
d\bar{\gamma}^2=-Bdv^2+2drdv+r^2d\omega^2_{n-1}~~~,
\ee
is the induced geometry on the brane. It
is easy to check that such a surface is geodesic and hence is a
solution of the Nambu-Goto equations (\ref{1}). 

Denote by $\xi^{\mu}$ a Killing vector field and let $J^b$ be a
conserved quantity corresponding to $\xi^{\mu}$. For a time-like
Killing vector $J^b=-E^b$, where $E^b$ is the energy, while for a
generator of rotations, $J^b$ is the angular momentum of the brane.
The flux per unit time $v$ of $J^b$ through the surface
$r$=const of the brane is 
\be\n{8a}
\dot{J}^{b}\equiv {dJ^b\over dv}=\int_{r=\mbox{const}}\sqrt{-g} 
T^{r\nu}\xi_{\nu}d^{N-1} \Omega
\ee
Due to the conservation law $T^{\mu\nu}_{~~~;\nu}=0$ the flux
$\dot{J}^{b}$ does not depend on $r$. Let us denote by $\dot{J}$ the
rate of the loss of the energy or angular momentum of the black hole
in this process, $\dot{J}=-\dot{J}^{b}$. By using  (\ref{3}),
(\ref{8a}) one finds
\be\n{8}
\dot{J}=\sigma\int_{r=\mbox{const}}d^{n-1}\omega
\sqrt{-\gamma}\gamma^{ab}\, X^{\mu}_{,a}n_{\mu}\,
X^{\nu}_{,b}\xi_{\nu}\, .
\ee
where $n_{\mu}=r_{,\mu}$.
The integral is taken over $(n-1)$
dimensional sphere and $d^{n-1}\omega$ is a measure on a unit sphere
$S^{n-1}$. 

For a static black hole, since both of the Killing vectors
$\partial_t$ and $\partial_{\phi}$ are tangent to the 
surface $r$=const, the energy and angular momentum of the black hole
are constant.

If a black hole is slowly rotating    the brane  changes its 
position from $\bar{X}^\mu(\zeta^a)$ to
$X^\mu(\zeta^a)=\bar{X}^\mu(\zeta^a) +\delta X^\mu(\zeta^a)$. The
brane perturbation can be found as a solution of  the linearized 
Nambu-Goto equations (\ref{1})
\begin{equation}\label{1a}
^{(n+1)}\bar{\Delta}\delta X^\mu+\delta ({}^{(n+1)}\Delta)\bar{X}^\mu
+2\bar{\gamma}^{ab}\bar{\Gamma}^\mu_{\nu\lambda}\bar{X}^\lambda_{,a}
\delta X^\nu_{,b}+
\delta\gamma^{ab}\bar{\Gamma}^\mu_{\nu\lambda}\bar{X}^\lambda_{,a}
\bar{X}^\nu_{,b}+
\bar{\gamma}^{ab}\delta\Gamma^\mu_{\nu\lambda}\bar{X}^\lambda_{,a}
\bar{X}^\nu_{,b}=0~~.
\end{equation}
Here all the quantities with the bar correspond to the static
black hole. Variations of the connections
$\delta\Gamma^\mu_{\nu\lambda}$ are due to the difference 
$\delta g_{\mu\nu}$ between
the metric of static and rotating black hole, 
$\delta g_{\mu\nu}=g_{\mu\nu}-\bar{g}_{\mu\nu}$. The variations
$\delta\gamma_{ab}$ 
of the metric induced on the brane world-sheet are caused
by variations $\delta g_{\mu\nu}$ of the bulk metric and changes
in the brane position
\begin{equation}\label{1b}
\delta\gamma_{ab}=\delta g_{\mu\nu}\bar{X}^\mu_{,a}\bar{X}^\nu_{,b}
+(\delta X_{,a},\bar{X}_{,b})+(\bar{X}_{,a},\delta X_{,b})~~,
\end{equation}
where we denote $(A,B)=\bar{g}_{\mu\nu}A^\mu B^\nu$, the inner product
with respect to the static metric.
Variation $\delta ({}^{(n+1)}\Delta)$ in (\ref{1a}) is the variation
of the box operator under the change of the brane metric.

In what follows we shall use the same coordinates
$\zeta^a=(v,r,\omega^{\alpha})$ on the perturbed brane so that the
coordinate volume $d^{n-1}\omega$ in (\ref{8}) remains the same. To
obtain the flux for a perturbed case one needs to calculate only 
variations $\delta X^{\mu}$ and $\delta\gamma_{a b}$ induced by  the
perturbations.

\section{A Kerr black hole interacting with a thin domain wall}
\setcounter{equation}0

\subsection{A solution of the perturbed equation}

We first consider a slowly rotating Kerr back hole interacting with a
thin domain wall. For a slowly rotating black hole the rotation
parameter $a$ is much smaller that the black hole mass $M$. In this
limit the Kerr metric (\ref{kerr})  takes the form
\be\n{19}
ds^2=d\bar{s}^2-2a\sin^2\theta \, d\varphi \left({r_0 \over r}\,
dv+dr\right)\, .
\ee
Here $d\bar{s}^2$ is the Schwarzschild  metric
\be\n{20}
d\bar{s}^2=-Bdv^2+2drdv+r^2(\sin^2\theta d\varphi^2+d\theta^2)\, ,
\ee
and $B=1-r_0/r$, $r_0=2M$.

Denote by $\alpha$ an angle between the axis of rotation and the
brane, then the equation of the unperturbed domain wall is
$\varphi=\bar{\varphi}(\theta)$ where
\be\n{21}
\sin\bar{\varphi}=\tan\alpha\, \cot\theta\, ,
\ee
and $\alpha \leq \theta \leq \pi-\alpha$ for $0\leq \alpha \leq
\pi/2$. 
The induced metric on the world-sheet of such a tilted domain
wall is
\be\n{22}
d\bar{\gamma}^2=-Bdv^2+2drdv+{r^2\sin^2\theta \over \sin^2\theta -
\sin^2\alpha}  d\theta^2\, .
\ee
One can check in this case the Nambu-Goto equations  (\ref{1}) reduce
to one equation
\be\n{86}
\bar{\varphi}_{,\theta\theta} +
2\cot\theta \bar{\varphi}_{,\theta}+\sin\theta\cos\theta 
\bar{\varphi}_{,\theta}^3=0~~
\ee
and (\ref{21}) is its solution.

If a black hole is slowly rotating    the position of the domain wall
becomes $X^\mu=\bar{X}^\mu+\delta X^\mu$. Those variations $\delta
X^\mu$ which are tangent to the brane can be gauged away. We use the
same coordinates $v$, $r$, and $\theta$ as in the unperturbed case to
parameterize the domain wall
\be\n{87}
v=\hat{v}\, ,\hspace{0.3cm}
r=\hat{r}\, ,\hspace{0.3cm}
\theta=\hat{\theta}\, ,
\end{equation}
so that the only non-trivial equation specifying a position of the
perturbed domain wall is
\be\n{88}
\varphi=\bar{\varphi}(\hat{\theta})+a\psi(\hat{r},\hat{\theta})\, .
\ee
One can check that for this choice of the gauge,
the corresponding $v,r$ and $\theta$ components of the Nambu-Goto 
equations (\ref{1a}) are satisfied identically.  
The equation for $\psi$ can be found from
the $\varphi$ component of (\ref{1a})
\begin{equation}\label{91}
^{(3)}\bar{\Delta} \psi+
2\bar{\gamma}^{ab}\bar{\Gamma}^\varphi_{\nu\varphi}\bar{X}^\nu_{,a}
\psi_{,b} +a^{-1}\left[ \bar{\gamma}^{ab}\delta\Gamma^{\varphi}_{\nu
 \lambda
}\bar{X}^\nu_{,a}\bar{X}^\lambda _{,b} +(\delta
^{(3)}\Delta)\bar{\varphi}+
\delta\gamma^{ab}~\bar{\Gamma}^{\varphi}_{\nu \lambda }\bar{X}^\nu
_{,a}\bar{X}^\lambda _{,b} \right]=0~~.
\end{equation}
The non-vanishing changes of the  metric induced on the
brane are
\begin{equation}\label{92}
\delta{\gamma}_{v\theta}=-a{r_0 \over r}
\sin^2\theta\bar{\varphi}_{,\theta}~~,~~
\delta{\gamma}_{r\theta}=a\sin^2\theta(-1+r^2 \psi_{,r})
\bar{\varphi}_{,\theta}~~,~~
\delta{\gamma}_{\theta\theta}=2ar^2\sin^2\theta \psi_{,\theta}
\bar{\varphi}_{,\theta}~~.
\end{equation}
The calculations of variations which enter (\ref{91}) are
straightforward but rather long. The calculations give the following
equation for  $\psi$
\be\n{93}
{}^{(3)}\bar{\Delta }\psi+
{2 \over r^2}\cot\theta \psi_{,\theta}+2{B \over r}\psi_{,r}= {1 \over
 r^3}
~~.
\ee
If $\psi$ is a function of $r$ only, this equation takes the
form
\be
\partial_r
(r B \psi')+{2B }\psi'={1 \over r^2}\, .
\ee
Its solution which is regular at the horizon and at infinity is
\be\n{94}
\psi=-{1\over r}~\, .
\ee
We demonstrate now that this is the only regular solution of
(\ref{93}).

\subsection{Uniqueness of a regular solution}

A solution of (\ref{93}) is a sum of (\ref{94}) and a solution of the
homogeneous equation 
\be\n{hom}
^{(3)}\bar{\Delta} \tilde{\psi}+
{2 \over r^2}\cot\theta \tilde{\psi}_{,\theta}+2{B \over r}
\tilde{\psi}_{,r}=0~~.
\ee
We show that it does not have non-trivial solutions
which are regular both at the horizon and infinity. 

Note that this equation allows a separation of variables. By writing
a solution in the form  $\tilde{\psi}(r,\theta)=\eta(r)\chi(\theta)$
and substituting it into the equation (\ref{hom}) one obtains
\be\n{rad}
{d\over dr}\left( r^3 B {d\eta\over dr}\right)=\Lambda r\eta\, ,
\ee
\be\n{ang}
{d\over d\theta}\left( {\Psi^3\over \sin\theta} {d\chi\over
d\theta} \right)=-\Lambda \sin\theta \Psi \chi \, .
\ee
Here $\Lambda$ is a separation constant  and
\be
\Psi=\sqrt{\sin^2\theta-\sin^2\alpha}\, .
\ee

At the boundary points of the interval $(\alpha,\pi-\alpha)$ the
function $\Psi$ vanishes, and  the angular equation (\ref{ang}) has
singular points. For a regular solution, $\chi$ remains finite at
these boundary points. The corresponding boundary value problem
determines the spectrum $\Lambda$. By multiplying the angular
equation (\ref{ang}) by $\chi$, integrating it over the interval
$(\alpha,\pi-\alpha)$, and making the integration by parts one
obtains
\be
\int_{\alpha}^{\pi-\alpha}{d\theta\over\sin\theta}\,
\Psi^3\left( {d\chi\over d\theta}\right)^2=
\Lambda \int_{\alpha}^{\pi-\alpha}\, d\theta\sin\theta \Psi\chi^2\, .
\ee
This relation implies that the separation constant $\Lambda$ is
non-negative.

Consider now solutions of the radial equation (\ref{rad}).
It is easy to see that these solutions at large $r$ have 
the following asymptotic:
$\eta(r)\sim r^\alpha$, where $\alpha(\alpha+2)=\Lambda$. Since
$\Lambda\ge 0$ one has either $\alpha\ge 0$ or $\alpha\le -2$. In the
first case the only solution which remains finite at infinity is
$\tilde{\psi}=$const. This solution is trivial and can be absorbed
by a change of the position of the unperturbed domain wall
$\varphi\to\varphi+$const. For all other solutions regular at infinity
one has $\alpha \le -2$. The equality takes place only for
$\Lambda=0$.

By multiplying the radial equation (\ref{rad}) by $\eta$, integrating
it over the interval $(r_0,\infty)$, and making the integration by
parts one obtains
\be
\Lambda \int_{r_0}^{\infty} dr r \eta^2+\int_{r_0}^{\infty} dr r^3 B
\left({d\eta\over dr}\right)^2= \left. r^3 B\eta{d\eta\over
 dr}\right|_{r_0}^{\infty}\, .
\ee
The expression in the right hand side vanishes both at the horizon
(where $B=0$) and at the infinity. Both integrals in
the left hand side are non-negative. The equality is possible only if
$\eta=0$. This proves that the homogeneous equation (\ref{hom}) does not
have non-trivial solutions. 

This result allows also the following interpretation. In the
unperturbed static space the induced metric on the domain wall is the
metric of a static (2+1)-dimensional black hole. The uniqueness
theorem we proved can be interpreted as a statement that this
(2+1)-dimensional black hole does not have `hairs' of a scalar field
$\tilde{\psi}$.

Thus (\ref{94}) is a unique regular solution. The result of our
computations for the vector $\delta X^\mu$ which describes
deformation of the position of the domain wall can be represented in a
more geometrical form. Let  $\xi_{(\phi)}=\partial_\varphi$ be the
Killing vector  of the Kerr metric related to rotation. Then, our
solution simply is 
\be
\delta X^\mu=-{a\over r} \xi_{(\phi)}^\mu\, .
\ee

\subsection{Angular momentum flux}

We show now that the rotation  induces the angular momentum flux
which is transfered from the black hole to the domain wall. To
calculate this flux we use the relation (\ref{8a}) which in the case
of a domain wall is
\be
\dot{J}^{b}\equiv {dJ^b\over dv}=\int_{r=\mbox{const}}\sqrt{-g} 
T^{r}_{\, \, \varphi}d^2 \Omega\, .
\ee
For the regular solution (\ref{94}) 
\be\n{24}
\sqrt{-g} T^{r}_{\, \varphi}=
\sigma r_0\,a \sin\theta\sqrt{\sin^2\theta-\sin^2\alpha}
\delta(\varphi-\bar{\varphi}(\theta)) 
\, . 
\ee
The rate of change of the angular momentum of the black hole $J=aM$ is
\be\n{25}
\dot{J}=-\pi \sigma a r_0\cos^2\alpha=-2\pi G_4 \sigma \cos^2\alpha J\, .
\ee
The angular momentum flux vanishes when the domain wall is in the
equatorial plane of the rotating black hole  \footnote{It is easy to
check that for an arbitrary value of $a$ the equatorial plane of the
Kerr black hole is a geodesic surface  and, hence, it obeys 
Nambu-Goto equations.}.  This is the final axisymmetric stationary
equilibrium configuration of the rotating black hole in the presence
of the domain wall.  The relaxation time when the black hole reaches
this final state is $T\sim (\pi G_4 \sigma \cos ^2\alpha)^{-1}$.

In Appendix we present similar results for a rotating Kerr black hole
attached to a cosmic string. Since Nambu-Goto equations for a test
stationary string allow an exact solution for an arbitrary value of
the rotation parameter, it is possible to obtain an exact expression
for the angular momentum flux beyond the slow rotation approximation.
The flux of the angular momentum vanishes only if the string is
directed along the axis of rotation of the black hole so that the
final state is axisymmetric.

\section{Interaction of higher dimensional rotating black
holes with branes}

\setcounter{equation}0

\subsection{Myers-Perry metric}

Now we consider 
a general case. We assume that a $N$-dimensional
rotating black hole is attached to a $n$-dimensional brane. If the
black hole size is much smaller that the size of extra dimensions,
and the tension of the brane is small, the gravitational field of the
black hole is described by the Myers-Perry (MP) metric
\cite{MP}. This metric beside the time-like at infinity Killing
vector $\xi_{(t)}^{\mu}$ has $[N/2]$ (the integer part of $N/2$)
mutually commutative and mutually orthogonal space-like Killing vectors
$\xi_{(i)}^{\mu}$ singled out by the property that they have closed
integral lines.   
The Killing vectors $\xi_{(i)}$ are elements of the Cartan
sub-algebra of the group of rotations $SO(n)$.
The MP metric is characterized by the gravitational
radius $r_+$ and by $[N/2]$  rotation parameters $a_i$. Such a black
hole has angular velocities $\Omega_i=a_i/(r_+^2+a_i^2)$. The vector
\be\n{null}
\eta=\xi_{(t)}+\sum_i \Omega_i \xi_{(i)}
\ee 
at the horizon becomes the null generator of the
horizon. (The summation over $i$ is performed from $i=1$ to
$i=[N/2]$.)

The metric takes slightly different form for even and odd $N$. For
{\em even} $N$ the MP metric in the ingoing coordinates is
\be\n{even}
ds^2=-dv^2+2dv\, dr +\sum_{i=1}^{N/2}\left[
(r^2+a_i^2)(d\mu_i^2+\mu_i^2\, d\varphi_i^2)-2a_i\mu_i^2\,dr\,
d\varphi_i +{mr^2\over \Pi\, F}(dv-a_i\, \mu_i^2\, d\varphi_i)^2
\right]\, ,
\ee
where
\be\n{PF}
\Pi=\prod_{i=1}^{[N/2]}(r^2+a_i^2)\, ,
\hspace{0.5cm}
F=1-\sum_{i=1}^{[N/2]}{a_i^2\mu_i^2\over r^2+a_i^2}\, .
\ee
The $\mu_i$ are coordinates on a unit $[N/2]$-dimensional sphere
\be
\sum_{i=1}^{[N/2]}\mu_i^2=1\, .
\ee
The $\varphi_i$'s are angles specifying the direction in the
bi-planes of rotation. The Killing vectors of MP metric are 
$\xi_{(i)}=\partial_{\varphi_i}$  The parameter $m=r_0^{N-2}$ is
related to the mass $M$ of the black hole, see (\ref{mass}). The
angular momenta $J_i$ of the black hole are defined as
\be\n{28}
J_i={{\cal A}_{N-1}\over 8\pi G^{N+1}} ma_i={2\over N-1} Ma_i\, .
\ee
The position of the event horizon $r_+$  is defined by the equation
\be
\Pi -mr^2=0\, .
\ee

The solution for {\em odd} $N$ is\footnote{It should be noted that the
sign of $a_i$ in both of the metrics (\ref{even}) and (\ref{odd}) 
is opposite to the sign adopted in the paper \cite{MP}. For our
choice of signs the metric (\ref{odd}) for $N=3$ coincides with Kerr
metric (\ref{kerr}).}
\[
ds^2=-dv^2+2dv\, dr +r^2\, d\mu^2+\sum_{i=1}^{[N/2]}\left[
(r^2+a_i^2)(d\mu_i^2+\mu_i^2\, d\varphi_i^2)
\right.
\]
\be\n{odd}
\left.
-2a_i\mu_i^2\,dr\,
d\varphi_i +{mr^2\over \Pi\, F}(dv-a_i\, \mu_i^2\, d\varphi_i)^2
\right]\, ,
\ee
where $\Pi$ and $F$ are defined by (\ref{PF}) and $\mu$ and $\mu_i$
obey the following relation
\be
\mu^2+\sum_{i=1}^{[N/2]}\mu_i^2=1\, .
\ee
The mass and angular momenta are defined as for the even $N$ case,
while a location of the event horizon is determined by the equation
\be
\Pi-mr=0\, .
\ee
In the absence of rotation both metrics (\ref{even}) and (\ref{odd})
reduce to the Tangherlini metric (\ref{4}). 

\subsection{Slow rotation limit}

Our aim is to investigate interaction of  branes with slowly rotating
black holes. For slow rotation, $a_i/r_0\ll 1$,  both  metrics
(\ref{even}) and (\ref{odd}) can be written  in the form  similar to
(\ref{19})
\begin{equation}\label{52}
ds^2=d\bar{s}^2-2\left[dr+\left({r_0\over r}\right)^{N-2}dv\right]
\sum_{i=1}^{[N/2]}a_i\mu_i^2 d\varphi_i~~.
\end{equation}
Here $d\bar{s}^2$ is the unperturbed metric
(\ref{4}). The metric on the unit sphere $d\Omega_{N-1}^2$ in (\ref{4}) 
has the form
\begin{equation}\label{48}
d\Omega^2_{N-1}=\sum_{i=1}^{[N/2]}(d\mu_i^2+\mu_i^2 d\varphi_i^2)~\, ,~~~
\mbox{for $N$ even}\, ,
\end{equation}
\begin{equation}\label{51}
d\Omega^2_{N-1}=d\mu^2+
\sum_{i=1}^{[N/2]}(d\mu_i^2+\mu_i^2 d\varphi_i^2)~\, ,~~~
\mbox{for $N$ odd}\, .
\end{equation}
By taking into account (\ref{4}), (\ref{48}), (\ref{51})
one can represent (\ref{52}) as
\be\n{27} 
ds^2=d\bar{s}^2- {2\over r^2}
\left[dr+\left({r_0\over r}\right)^{N-2}dv\right] \varrho_{\mu}dx^{\mu}\, ,
\ee
\be\n{varrho}
\varrho_{\mu}=\bar{g}_{\mu\nu}\varrho^{\nu}\, ,\hspace{0.5cm}
\varrho^{\mu}=\sum_i a_i \xi_{(i)}^{\mu}\, .
\ee
In the linear approximation $r_+\approx r_0$, $\Omega_i\approx
a_i/r_0^2$, and instead of (\ref{null}) one has 
\be\n{null2}
\eta =\xi_{(t)}+r_0^{-2}\varrho~~.
\ee
Therefore, in the linear approximation $\varrho$ is related
to the vector $\eta$ which at the horizon becomes the null
generator of the horizon surface.

\subsection{Solution to linearized Nambu-Goto equations}

Let us consider a brane interacting with a higher-dimensional
slowly rotating  black hole described by metric (\ref{52}). As we
have already pointed out the rotation results in the deformation of
the brane world-sheet described by the vector field $\delta X^\mu$. 
This deformation  is the solution to  linearized Nambu-Goto equations
(\ref{1a}). We demonstrate now that these equations
allow a solution of the form $\delta X^\mu=\psi \,\varrho^\mu$ where
$\psi=\psi(r)$ and $\varrho^\mu$ is the Killing vector field
(\ref{varrho}). It is this form of solution which we found for the 
domain wall in case of Kerr black hole.

By using (\ref{52})  one can show that in coordinates  where
$\xi_{(i)}^\mu=\delta^\mu_{\varphi_i}$ the following relations are
valid
\[
^{(n+1)}\bar{\Delta}\delta X^\mu={1 \over r^{n-1}}\partial_r
(r^{n-1} B \psi')\varrho^\mu~~,
\]
\[
(\delta ^{(n+1)}\Delta)\bar{X}^\mu=
{1 \over r^{n-1}}\partial_r
\left(r^{n-1}\left({1 \over r^2}-B\psi'\right)\right)
(\varrho^{\parallel})^\mu~~,
\]
\[
2\bar{\gamma}^{ab}\bar{\Gamma}^\mu_{\nu\lambda}\bar{X}^\lambda_{,a}
\delta X^\nu_{,b}={2B \over r}\psi' \varrho^\mu~~,
\]
\[
\delta\gamma^{ab}\bar{\Gamma}^\mu_{\nu\lambda}\bar{X}^\lambda_{,a}
\bar{X}^\nu_{,b}={2 \over r}\left({1 \over r^2}-B\psi'\right)
(\varrho^{\parallel})^\mu~~,
\]
\[
\bar{\gamma}^{ab}\delta\Gamma^\mu_{\nu\lambda}\bar{X}^\lambda_{,a}
\bar{X}^\nu_{,b}=-{n-1 \over r^3}\varrho^\mu~~.
\]
Here $\varrho^{\parallel}$ is the projection of the vector 
$\varrho$ on the plane tangent to the brane. 
With these relations linearized Nambo-Goto
equations (\ref{1a}) take the simple form
\be\n{1cc}
\left[{1 \over r^{n-1}}\partial_r
(r^{n-1} B \psi')+{2B \over r}\psi'-{n-1 \over r^3}\right]
(\varrho^{\perp})^\mu=0~~~,
\ee
where $\varrho^{\perp}=\varrho-\varrho^{\parallel}$ is the 
projection of $\varrho$ in the hyperplane orthogonal to the brane. As
follows from (\ref{1cc}), if $\varrho^{\perp}=0$ the Nambu-Goto
equations are satisfied automatically. In this case a deformation
vector $\delta X^\mu$ is tangent to the brane world sheet. Such
deformations can be gauged away. If $\varrho^{\perp}\neq 0$  the
square bracket in (\ref{1cc}) must vanish. This condition yields
equation for function $\psi$, which has a single solution regular
both at the horizon $r=r_0$ and at infinity
\be\n{perx}
\psi(r)=-\int^{\infty}_r {dr \over r^2B}\left[1-\left({r_0 \over
r}\right)^{n-1}\right]~~.
\ee
For a domain wall interacting with a Kerr black hole ($n=2$) this
expression reduces to the earlier obtained result $\psi=-1/r$. In
case of cosmic strings, $n=1$, the only regular solution is $\psi=0$
for any dimension $N$ of the bulk black hole.

A general solution of the linearized Nambu-Goto equations (\ref{1a})
is a sum of the obtained solution $\psi(r) \,\varrho^\mu$ and a
solution of the homogeneous equations.  In the case of the domain
wall we proved that there is no non-trivial homogeneous solutions
which are regular both at the event horizon and infinity.  One can
expect that in the higher dimensional case  the situation is similar.
One can check that a trivial homogeneous solution is $\sum_i^{[N/2]}
c_i \xi_{(i)}^\mu$ ($c_i$ are  constants). It can be eliminated by
appropriate shifts of coordinates $\varphi_i$.

\subsection{Final stationary state of a rotating black hole}

We can now describe a final stationary state which is reached
by a black hole as a result of the interaction with the brane.
To this aim we investigate loss of the 
angular momentum of the black hole. Since for the static metric
$d\bar{s}^2$ $\dot{J}_i$
vanishes it is sufficient to calculate the variation of (\ref{8})
induced by the metric perturbations. We have
\be\n{29}
\dot{J}_i=\sigma\int_{r=\mbox{const}}d^{n-1}\omega
\delta(\sqrt{-\gamma}\gamma^{ab}\, X^{\mu}_{,a}n_{\mu}\,
X^{\nu}_{,b}\xi_{(i)\nu})\, .
\ee
It is easy to check that in the linear in $a_i$ approximation the
following relations for the variations induced by the perturbed metric
(\ref{27}) are valid: $\delta(\sqrt{-\gamma})=0$,
\be\n{a}
\xi_{(i) a}\delta\gamma^{ra}= {1\over r^2}\, 
(\xi^{\parallel}_{(i)},\varrho^{\parallel})-B(\xi^{\parallel}_{(i)},
\delta X_{,r}^{\parallel})\, ,
\ee
\be
\gamma^{ra}\delta\xi_{(i)a}=- {a_i\over r^2}\,  (\xi_{(i)})^2\, ,
\ee
\be\n{b}
\gamma^{ra}\, \xi_{(i) \lambda} \delta X^{\lambda}_{,a}=B
(\xi_{(i)},\delta X_{,r})\, .
\ee
As earlier we denote by $p^{\parallel}$ a projection of the vector $p$ on the
brane.
$(p,q)$ is a scalar product of vectors $p$ and $q$ in the
unperturbed metric, $(p,q)=\bar{g}_{\mu\nu}p^{\mu}q^{\nu}$.

The flux of the angular momenta from the black hole to the brane
changes the angular momenta of the black hole (\ref{28}). In the
linear approximation the equations for the
change of the angular momenta of the black hole can be written as
follows
\be\n{dyn}
\dot{{\bf J}}=-{\bf K}\, {\bf J}\, .
\ee
We use bold-faced quantities for vectors and tensors in the space of
rotation parameters, so that  ${\bf J}$ and  ${\bf K}$ have
components  $J_i$,  and $K_{ij}$.  Relations (\ref{28}),
(\ref{perx})--(\ref{b}) enable one to get ${\bf K}$  in the form
\be\n{K}
K_{ij}=(N-1)\sigma r_0^{n-1}k_{ij}/(2M)\, ,
\ee 
\be\n{int}
k_{ij}=\int_{S^{n-1}} d\omega^{n-1}\, 
{1\over r^2}\, (\xi^{\perp}_{(i)},\xi^{\perp}_{(j)})\, .
\ee
We denote by $p^{\perp}$  a projection of a vector $p$  orthogonal to
the brane. In an agreement with the conservation law, the angular
momentum flux does not depend on the radius $r$ of the surface where
it is calculated. 

Note that in the linear in $a_i$ approximation
$\dot{M}=\dot{r}_0=0$ and $M$ and $r_0$ in (\ref{K}) are considered
as constant parameters. The evolution equation (\ref{28}) can also be
written as $\dot{{\bf a}} =-{\bf K}\, {\bf a}$, where $\bf a$ is a
vector with components $a_i$. 
This equation shows
that  the black hole can
be stationary, $\dot{\bf a}=0$, if and only if ${\bf a}$ is the zero
vector, ${\bf K a}=0$. In this case the equation ${\bf a}^{T}{\bf K
a}=0$ implies that
\be
\int_{S^{n-1}} d\omega^{n-1}\, (\varrho^{\perp},\varrho^{\perp})=0\, ,
\ee
and hence $\varrho^{\perp}=0$. This means that the corresponding
Killing vector $\varrho$ generates transformations that preserve a
position of the brane. The stationary metric of the final black hole
configuration in this case is given by (\ref{27}) where $\varrho$ is
a  vector tangent to the brane.  Since in the linear approximation
$\varrho$ is related to  the null generator of the black hole
horizon, $\eta=\xi_{(t)}+r_0^{-2}\varrho$, in the final state of the
black hole  $\eta$ is tangent to the  brane.  

This property is valid for any value of a rotation parameter. In the
stationary case the surface area of the black hole horizon is
constant. By using the Raychaudhuri equation one has
$T_{\mu\nu}\eta^\mu\eta^\nu=0$, where $T_{\mu\nu}$ is the
stress-energy tensor of the brane. Since $T_{\mu\nu}$ is proportional
to the induced metric, the null generator of the horizon $\eta$ is
tangent to the brane on the horizon\footnote{A similar condition for
a static black hole was discussed   in \cite{EHM} in a slightly
different context.}. 

However, the result we have found for slowly rotating black holes is
more than that. We have shown that vector fields $\eta$ and 
$\varrho$ are tangent everywhere to the brane-world-sheet, not only at
the horizon.  Moreover, because the deformation of the brane has the
form $\delta X^\mu=\psi(r) \varrho^\mu$, it follows that in the
final  stationary state  $\delta X^{\perp \mu}=0$. Since the tangent
to the brane components of $\delta X^{\mu}$ can always be gauged
away,  in  the limit of slow rotation the final state of a stationary
brane is not deformed. It is easy to see that condition that $\varrho$
is tangent to the brane implies that the brane stress-energy tensor
obeys  the condition $\n{Lie} {\cal L}_\varrho T^{\mu\nu}=0$. This is
a result of the invariance of the brane under the symmetry
transformations generated by $\varrho$. In the $N=3$ case this is a
condition of the axial symmetry, in accordance with the Hawking
theorem \cite{Hawk:72}.

\section{Evolution equations for special cases}
\setcounter{equation}0

In a general case, the evolution in the space of the rotation
parameters and a number of free rotation parameters available in the
final state are determined by the properties of the matrix ${\bf k}$.
From a mathematical point of view this is a problem of mutual
orientation of a Cartan sub-algebra of fields $\xi_i$ related to the
rotation planes of the black hole, and a sphere $S^{n-1}\subset
S^{N-1}$ related to the brane position. To illustrate  different
possibilities we  consider some special cases.

We consider a unit sphere $S^{N-1}$ embedded in a Euclidean space
$R^N$ and  let $X^A$, $A=1,...,N$, be the Cartesian coordinates in
$R^N$.   A sphere $S^{n-1}$ representing the position of the brane
can be obtained as an intersection of an $n$-dimensional hyperplane
$P_n$ with $S^{N-1}$.  We choose the  coordinates $X^A$ so that the
plane $P_n$ is described by the  equations $X^{n+1}=..=X^N=0$. We
denote by $\vec{e}_A=\partial/ (\partial X^A)$ the basis vectors in
$R^N$. The first $n$ vectors span $P_n$,  while the last $N-n$
vectors of this basis are orthogonal to $P_n$. Any vector $\vec{v}$
in $R^N$ is a sum of its projections parallel ($\vec{v}^{\ \parallel}$)
and orthogonal ($\vec{v}^{\perp}$) to the hyperplane $P_n$.

The Cartan sub-algebra for $SO(N)$ can be specified by choosing
$[N/2]$ mutually orthogonal bi-planes.  Let $\vec{p}_i$ and
$\vec{q}_i$ be two orthonormal vectors which span the $i$-th
bi-plane of rotation. We denote by 
\be
(\vec{v}\cdot\vec{u})=\sum_{A=1}^{N} v_A u_A
\ee
a scalar product of two vectors  $\vec{v}$ and $\vec{u}$ in $R^N$. By
definition
\begin{equation}\label{110}
(\vec{p}_i\cdot \vec{p}_j)=(\vec{q}_i\cdot \vec{q}_j)=\delta_{ij}~~~,~~~~
(\vec{p}_i\cdot \vec{q}_j)=0~~~.
\end{equation}
A Killing vector field $\vec{\xi}_{(i)}$ generating a rotation in a
given bi-plane can be written as
\be\n{k-i}
\vec{\xi}_{(i)}=(\vec{p}_i\cdot \vec{X}) \vec{q}_i-(\vec{q}_i\cdot \vec{X}) \vec{p}_i~~.
\ee
Let us denote
\be
\vec{\zeta}_{(i)}=\left. \vec{\xi}^{\perp}_{(i)}\right|_{S^{n-1}}\, ,
\ee
then we have
\be\n{k-i-2}
\vec{\zeta}_{(i)}=(\vec{p}^{\,\, \parallel}_i\cdot \vec{X}) \vec{q}^{\perp}_i-
(\vec{q}^{\,\, \parallel}_i\cdot \vec{X}) \vec{p}^{\perp}_i~~.
\ee
Since $\vec{X}$ in this relation belongs to $S^{n-1}$, one has
$$
(\vec{X}\cdot \vec{X})=\sum_{A=1}^n X^{A} X^{A}=1~~.
$$
The scalar product which enters the matrix
$k_{ij}$, see (\ref{int}), is
\be\n{127}
{1 \over r^2}(\xi^{\perp}_{(i)},\xi^{\perp}_{(j)})=
(\vec{\zeta}_{(i)} \cdot \vec{\zeta}_{(j)})~~~.
\ee
After integrating over $S^{n-1}$ one finds
$$
k_{ij}=\int_{S^{n-1}} d\omega^{n-1}\,  (\vec{\zeta}_{(i)}\cdot
\vec{\zeta}_{(j)})=Z_n\, \left[ (\vec{q}^{\perp}_i\cdot
\vec{q}^{\perp}_j)(\vec{p}^{\,\, \parallel}_i\cdot \vec{p}^{\,\,
\parallel}_j) +\right.
$$
\be\n{128}
\left. (\vec{p}^{\perp}_i\cdot \vec{p}^{\perp}_j)(\vec{q}^{\,\,
\parallel}_i\cdot \vec{q}^{\,\, \parallel}_j)
-(\vec{q}^{\perp}_i\cdot \vec{p}^{\perp}_j)(\vec{p}^{\,\,
\parallel}_i\cdot \vec{q}^{\,\, \parallel}_j)
-(\vec{p}^{\perp}_i\cdot \vec{q}^{\perp}_j)(\vec{q}^{\,\,
\parallel}_i\cdot \vec{p}^{\,\, \parallel}_j) \right]\, .
\ee
Here we used the fact that 
\be
\int_{S^{n-1}}d\omega^{n-1} X^{A}\, X^{B}=\delta^{AB} Z_n\, ,
\hspace{0.5cm}
Z_n=
{\pi^{n/2} \over \Gamma\left({n+2 \over 2}\right)}~~. 
\ee
Formula (\ref{128}) enables one to infer properties of the
matrix $k_{ij}$.  Let us discuss some simple cases.

(1) Suppose the $i$-th bi-plane of rotation has the orientation for
which $\vec{p}_i^{\,\, \parallel}=\vec{q}_i^{\,\, \parallel}=0$, then
the components of ${\bf k}$ with a chosen index $i$ vanish. The
rotation in the given $i$-th bi-plane transforms the brane into
itself. There are no fluxes of the $i$-th component of the angular
momentum, $\dot{J}_i=0$. An example of this case is a cosmic string
directed along the rotation axis of the Kerr black hole.

(2) Suppose the $i$-th bi-plane has the orientation for which
$\vec{p}_i^{\perp}=\vec{q}_i^{\perp}=0$. Then again, there is no flux
of $J_i$, $\dot{J}_i=0$, and the rotation in the $i$-th bi-plane
transforms the brane into itself. An example of this situation is a
domain wall lying in the equatorial plane of the Kerr black hole.

(3) Suppose that $\vec{q}_i^{\,\, \parallel}=0$ but $\vec{p}_i^{\,\,
\parallel}\neq 0$. Then $(\vec{q}^{\perp}_i\cdot
\vec{q}^{\perp}_j)=(\vec{q}_i\cdot \vec{q}_j)=\delta_{ij}$,
$(\vec{q}^{\perp}_i\cdot \vec{p}^{\perp}_j)=(\vec{q}_i\cdot
\vec{p}_j)=0$ and 
\be\n{129}
k_{ij}= Z_n \delta_{ij}\sin^2\alpha_i \,~~,~~\mbox{for any}~~j~~,
\ee
where $\sin^2\alpha_i\equiv(\vec{p}^{\,\, \parallel}_i\cdot
\vec{p}^{\,\, \parallel}_i)$. The parameter $\alpha_i$ is related to
the angle  $\pi/2-\alpha_i$ between the brane  and the 
$i$-th bi-plane. This configuration is realized when a cosmic string
is tilted to the rotation axis of the Kerr black hole. In this case
$\alpha_i$ is the angle between the string and the axis. The flux  of
the angular  momentum is proportional to $\sin^2\alpha_i$, in
agreement with (\ref{17}).
 
(4) Suppose that $\vec{q}_i^{\perp}=0$ but $\vec{p}_i^{\perp}\neq 0$.
Then $(\vec{q}^{\,\, \parallel}_i\cdot \vec{q}^{\,\, \parallel}_j)=\delta_{ij}$,
$(\vec{q}^{\,\, \parallel}_i\cdot \vec{p}^{\,\, \parallel}_j)=0$
and 
\be\n{130}
k_{ij}=\delta_{ij}\cos^2\alpha_i Z_n~~,~~\mbox{for any}~~j~~.
\ee
For the domain wall tilted at the angle $\alpha_i$ to rotation axis
of the Kerr black hole  one gets the momentum flux proportional to
$\cos^2\alpha_i$, in agreement with (\ref{25}).

The above examples show that the  matrix $k_{ij}$ can have
eigenvectors with positive eigenvalues. These eigenvectors define the
directions in the space of parameters $a_i$ for which the evolution
is  damping. The  damping is caused by the `friction' which is a
result of the interaction between the black hole and the brane.

\section{Discussion}

Let us summarize the results obtained in this paper. We considered
interaction of branes with rotating higher dimensional black holes.
Such systems include several physically interesting examples, such as
cosmic strings and thin domain walls interacting with the Kerr black
hole, as well as rotating black holes in a space-time with large extra
dimensions. If a black hole does not rotate, in the zero order in the
string tension (test string approximation)  a static brane located in
the black-hole's `equatorial plane' is always a solution of the
Nambu-Goto equations. In the presence of rotation, a stationary brane
attached to a black hole is deformed. In a general case this
deformation  $\delta X^\mu=\psi(r)\varrho^\mu$ is determined by the
Killing vector $\varrho^\mu$, (\ref{varrho}), which describes the
change of the black hole metric under slow rotation, see (\ref{27}). 
For Kerr black hole we demonstrated  that these solutions obeying the
regularity conditions at the horizon and infinity are unique. One can
make a conjecture that this property is valid for arbitrary number of
space and brane dimensions.

We demonstrated that  in a general case there exists an angular
momentum transfer from a black hole to the attached brane until
$\varrho$ becomes tangent to the brane. The brane stress energy
tensor in this state obeys condition ${\cal
L}_{\varrho}T^{\mu\nu}=0$.   One may expect that there should exist a
generalization of this result for a rotating black hole surrounded
by   a stationary distribution of matter or field \cite{AF},  so that
a higher dimensional analogue of the Hawking theorem \cite{Hawk:72}
is valid. 

The flux of the angular momentum is determined by the
tension of the brane $\sigma$ and by the black hole radius $r_0$ 
\be\n{rel}
\dot{J}\sim -G\sigma r_0^{1-p} J\, , 
\ee 
where $p$ is the  number of brane co-dimensions dimensions. The
characteristic time of the relaxation process during which the black
hole reaches its final state can be estimated by using (\ref{rel}) 
\be\n{rel1}
T\sim r_0^{p-1}/ (G_{N+1}\sigma)\, \sim T_* ({r_0/ L_*})^{p-1}\,
(\sigma_*/\sigma)\, .
\ee
Here $p=N-n$ is the number of extra dimensions,
$\sigma_*=M_*/L_*^n$ and quantities 
$M_*$, $L_*$ are, respectively,  
the fundamental mass and the length of the theory.

It should be  noted that the black hole can also loose its bulk
components of the rotation by emitting  Hawking quanta in the bulk. 
The characteristic time of this process is $T_{H}\sim T_*
(r_0/L_*)^N$. For black holes which can be treated classically 
$r_0\gg L_*$, $T_H \gg T$. Thus the friction effect induced by the
brane is the dominant one.

We would like to conclude the paper by the following remark. We
focused our attention on the higher dimensional space-times  
with vanishing bulk cosmological constant (ADD model \cite{ADD}). 
A similar problem
concerning general properties of higher dimensional rotating black
holes with the horizon radius $r_{0}$ can be addressed  in
the Randall-Sundram (RS) models \cite{RS} provided the bulk 
cosmological constant is much smaller than $r_{0}^{-2}$. 
A characteristic property of
such models is the existence of $Z_2$ symmetry. Under
$Z_2$ transformation the brane remains unchanged, while the
components of any vector orthogonal to the brane change their sign. 
Thus $Z_2$ symmetry implies $\varrho^{\perp}=0$.  Hence a stationary
black hole attached to the brane in the RS-model can rotate only
within the brane.

The relaxation process related to the presence of $\varrho^{\perp}$
which is typical for the ADD-model is absent in the
RS-model. This is an additional signature which in principle may
allow one to distinguish between these models in observations.

\noindent
\section*{Acknowledgments}

\indent V.F. and D.F. kindly acknowledge the support from  the NATO
Collaborative Linkage Grant (979723). The work of V.F. and D.F. is
also partially supported   by the Killam Trust and the Natural
Sciences and Engineering Research Council of Canada.
DS thanks the DOE and the Michigan Center for Theoretical Physics for
support at the University of Michigan.

\appendix

\section{An interaction of a Kerr black hole with a cosmic string}
\setcounter{equation}0

In this appendix we consider an interaction of a Kerr black hole with
a cosmic string. This case is of special interest since the
stationary test string equations can be solved exactly for an
arbitrary value of the rotation parameter $a$. The Kerr metric in
the  Boyer and Lindquist coordinates  is
\be\label{kerr}
ds^2=-\left( 1-\frac{2Mr}{\Sigma
}\right) dt^2-\frac{4Mra\sin^2\theta }{\Sigma}\,dt\,d\phi +
\frac{\Sigma}{\Delta}\,dr^2+\Sigma \,d\theta ^2+
{A \sin^2\theta\over{\Sigma}} \,d\phi ^2 ,
\ee
where 
\[
\Sigma \equiv r^2+{a^2\cos^2\theta}\,, \hspace{1cm}
 \Delta \equiv r^2-{2Mr}+{a^2}\,,
\]
\be
A=\left(r^2+{a^2}\right)^2 -{a^2\Delta\sin ^2\theta}\, .
\ee
The event horizon is located at $r=r_+=M+\sqrt{M^2-a^2}$.

By making the following coordinate transformation
\be
dv=dt+(r^2+a^2)\,\frac{dr}{\Delta}\,,  \hspace{1cm} 
d\varphi=d\phi+a\,\frac{dr}{\Delta}\,. 
\ee
we obtain this metric in the
{\em Kerr ingoing coordinates}
\[ \n{kerrin}
ds^2=-\,\left[1-\Sigma^{-1}(2Mr)\right]\,dv^2+2\,dr\,dv+\Sigma\, d\theta^2
 +\,\Sigma^{-1}\left[(r^2+a^2)^2-\Delta\,
a^2\sin^2\theta\,\right]\sin^2\theta \,d\varphi^2
\]
\be
 -\,2a\sin^2\theta\,d\varphi\,dr-4a\,\Sigma^{-1}Mr\sin^2\theta
\,d\varphi\,dv\,.
\ee
These coordinates are regular on the future event horizon.
One also has $\sqrt{-g}=\Sigma\, \sin\theta$.
The Kerr metric possesses two  Killing vectors 
$\xi_{(t)}^{\mu}=\delta^{\mu}_{v}$ and
$\xi_{(\phi)}^{\mu}=\delta^{\mu}_{\varphi}$.

In what follows we shall consider stationary fluxes of  energy and
angular momentum of some distribution of matter through a surface
$r=$const
\be\n{24a}
\Delta E=\int T_{\mu}^{\, \nu}\, \xi_{(t)}^\mu d\sigma_\nu~~, 
\hspace{0.5cm}
\Delta J=-\int T_{\mu}^{\, \nu}\, \xi_{(\varphi)}^\mu d\sigma_\nu~~. 
\ee
Here $T_{\mu\nu}$ is the stress-energy tensor of the matter and 
\be
d\sigma_{\mu}=r_{,\mu}\sqrt{-g}\, dv\, d\theta\, d\varphi\, .
\ee

For a stationary configuration the (constant) rate of energy  and 
angular momentum fluxes from the black hole through the $r=$const
surface are
\be\n{dotE}
\dot{E}\equiv {dE\over dv}=-\int  d\theta d\varphi\, \sqrt{-g}\, T_{v}^{r}\, ,
\hspace{0.5cm}
\dot{J}\equiv {dJ\over dv}= -\int  d\theta d\varphi\, \sqrt{-g}\,T_{\varphi}^{r}\, .
\ee

If a part of initially infinite string is captured by a black hole its
world-sheet in the black hole exterior consists of two segments. Let us
consider a case when a cosmic string  is stationary. 
In \cite{FrHeLa} it was proved that any segment of this string   in
the Kerr space-time which starts at infinity and crosses the infinite
red-shift surface, remaining regular there,   has a world-sheet
generated by the Killing vector $\xi_{(t)}^{\mu}$ and a principal
null geodesic. The string segment regularly crosses the future
event horizon. In the incoming Kerr coordinates its equation has the
following simple form $\theta=$const and $\varphi=$const. Without
loss of generality we put $\varphi=0$, so that the string segment is
described by the equations
\be\n{12}
v=\hat{v}\, ,\hspace{0.3cm}
r=\hat{r}\, ,\hspace{0.3cm}
\theta=\alpha\, ,\hspace{0.3cm}
\varphi=0\, ,
\end{equation}
where $\hat{r},\hat{v}$ are coordinates on the string world-sheet. We
denote by   $\alpha$ the angle between the string segment and the
rotation axis as measured at infinity.  We consider a  configuration
where there is another  string segment attached to the black hole
with $\theta=\pi-\alpha$, $\varphi=\pi$. Because of the symmetry
$\theta\to\pi-\theta$ and $\phi\to \pi+\phi$ of the Kerr metric the
strings' tension for such a symmetric configuration does not result
in the accelerated motion of the black hole as a whole.

The metric induced on the string world-sheet is
\be\n{13}
\gamma_{ab}d\zeta^a d\zeta^b=-B dv^2+2dr dv~~,
\hspace{0.5cm}
B=1-{r_0r\over r^2+a^2\cos^2\alpha}\, . 
\end{equation}

For a stationary configuration the (constant) rate of 
angular momentum flux through the $r=$const surface is
\be\n{15}
\dot{J}\equiv {dJ\over dv}= -\int  d\theta d\varphi\,
\sqrt{-g}\,T_{\varphi}^{r}\, ,
\ee
where 
\be\n{16}
\sqrt{-g} T^{r}_{\varphi}=
\sigma a \sin^2\alpha \delta(\varphi)\delta(\theta-\alpha)~~. 
\ee
The angular momentum flux remain the
same for any $r$ and is given by the following relation
\be\n{17}
\dot{J}=-2\sigma a \sin^2\alpha=-{4\sigma  \sin^2\alpha\over r_0}\, J ~~. 
\ee
The coefficient $2$ appears since there are 2 string segments. 
$\dot{J}$ vanishes when the direction of the string segments coincides
with the symmetry axis, $\alpha=0$ or $\alpha=\pi$. 

This result implies that the black hole looses its angular momentum
whenever the string does not coincide with the axis of the rotation.
In other words, the final equilibrium configuration the black hole is
either non-rotating or its angular momentum is directed along the
string. The characteristic time of this process estimated from
(\ref{17}) is $T\sim~r_0/( 4\sigma  \sin^2\alpha)$.

\end{document}